\documentclass[aps,twocolumn,floatfix, showkeys, superscriptaddress, nofootinbib]{revtex4-1}
\usepackage{amsmath}
\usepackage{graphicx,bm,hhline, xcolor, float}
\usepackage[colorlinks=true]{hyperref}

\newcommand{\br}{{\bm r}}

\begin{document}
\title{Excited States in Warm and Hot Dense Matter}

\author{C. E. Starrett}
\email{starrett@lanl.gov}
\affiliation{Los Alamos National Laboratory, P.O. Box 1663, Los Alamos, NM 87545, U.S.A.}

\author{T. Q. Thelen}
\affiliation{Los Alamos National Laboratory, P.O. Box 1663, Los Alamos, NM 87545, U.S.A.}

\author{C. J. Fontes}
\affiliation{Los Alamos National Laboratory, P.O. Box 1663, Los Alamos, NM 87545, U.S.A.}

\author{D. A. Rehn}
\affiliation{Los Alamos National Laboratory, P.O. Box 1663, Los Alamos, NM 87545, U.S.A.}

\date{\today}
\begin{abstract}
Accurate modeling of warm and hot dense matter is challenging in part due to the multitude of excited states that must be considered.  In thermal density functional theory, these excited states are averaged over to produce a single, averaged, thermal ground state.  Here we present a variational framework and model that includes explicit excited states.  In this framework an excited state is defined by a set of effective one-electron occupation factors and the corresponding energy is defined by the effective one-body energy with an exchange and correlation term.  The variational framework is applied to an atom-in-plasma model (a generalization of the so-called average atom model).  Comparisons with a density functional theory based average atom model generally reveal good agreement in the calculated pressure, but the new model also gives access to the excitation energies and charge state distributions.  
\end{abstract}
\maketitle

\section{Introduction}
In warm and hot dense matter the electronic structure comprises a complex and very large set of excited states of the Hamiltonian.  Electrons are excited to populate these states through collisions and absorption of electrons and photons.  The resulting properties of the plasma, such as the equation of state (EOS), opacity, and transport coefficients, can, in principle, be calculated by taking appropriate averages over these excited states.  Experiments can probe in detail these excited states \cite{ciricosta2012direct, nagayama19, bailey15, kritcher2020measurement, fletcher15, hollebon19, vinko2020time}, providing stringent tests of our models.

There are many approaches to modeling warm and hot dense matter.  Broadly speaking, at lower densities, a particularly successful approach is to start with the atomic structure of the isolated atoms or ions, and then to correct for plasma effects \cite{hakel06, colgan16, iglesias1996updated, Stewart66lowering}.  Typically, these models become less successful at higher densities where a consistent treatment of plasma effects becomes crucial.  At high densities, notable methods include path-integral Monte-Carlo (PIMC) \cite{militzer15, pollock1984simulation} and density functional theory (DFT) based simulations \cite{collins95, hu2011first, holst2008thermophysical, starrett20ms}.    PIMC is so far largely limited to EOS applications and lower $Z$ materials.  The widely popular DFT \cite{hohenberg}, through its finite temperature extension \cite{mermin65}, has been applied to a wide variety of properties and is generally very successful.  

However, attempts to apply DFT to the opacity of moderately or non-degenerate systems reveals a fundamental weakness of the approach \cite{shaffer22dense, karasiev2022first, gill2021time}.  As shown in these references, the calculated opacities differ wildly from the measured data.  In DFT, it is the properties of the (thermal) ground state of the system that are calculated.  To calculate excited state properties, time-dependent (TD-) DFT~\cite{zangwill80, runge84} or ensemble DFT (EDFT)~\cite{theophilou1979energy,gross1988rayleigh,gross1988density,oliveira1988density} is required. 

TD-DFT calculations performed using the adiabatic local density approximation (ALDA)~\cite{ullrich2011time} fail dramatically when applied to these opacity problems~\cite{gill2021time,ullrich2014time}. Although TD-DFT in principle is exact and allows for the calculation of all excited state properties, the ALDA uses only the instantaneous density for evaluation of the exchange-correlation functional. Multiple excitations are known to arise from nonadiabatic electron correlation effects that are neglected in the ALDA. This indicates that to properly treat multiple excitations, memory and dissipation effects need to be included explicitly in the exchange-correlation functional~\cite{maitra2004double,ullrich2006time,sangalli2011double,elliott2011perspectives}. Hence, despite being in principle exact, in practice for these problems, the standard approximations (e.g., ALDA) are grossly inadequate.

The EDFT approach is an alternative to TD-DFT for calculating excited states, and is perhaps more naturally suited for inclusion of multiple excitations. This approach, formulated via a variational principle for ensembles of excited states, provides a modified Hohenberg-Kohn theorem that establishes an exact mapping between the potential and ensemble density. The benefit of this approach is that, via a corresponding Kohn-Sham scheme, the density for an ensemble of multiple excited states can be solved for simultaneously, and the energies of the excited states can be obtained. While this approach has that particular advantage, it is less frequently used than TD-DFT and many open questions remain, such as how to construct exchange-correlation functionals that go beyond the ``quasi-LDA''~\cite{kohn1986density} and how to address the issue of ghost states~\cite{gidopoulos2002spurious}, which are in some ways analogous to electron self-interaction in ground-state DFT. Furthermore, temperature effects are not addressed in this approach. The approach we take in this paper is more similar to EDFT than to TD-DFT, though it is not formulated on a rigorous theoretical foundation, but instead more closely follows atomic physics approaches for calculating excited states. However, unlike EDFT, temperature is included explicitly in our approach.

In DFT the energy is divided into that of a non-interacting system (that gives the same electron density as the full interacting system) and a correction (the exchange and correlation term).  Currently, the most accurate numerical scheme uses the Kohn-Sham (KS) approach \cite{kohn1965self}.  In KS-DFT, the non-interacting energy is found by solving a single-particle Schr\"odinger equation for an effective one-body interaction potential, determined by minimising the energy with respect to the density.  This solution gives the eigenstates of this effective Hamiltonian and the electronic density is constructed by filling these states according to Fermi-Dirac statistics.  Even at elevated temperatures this scheme is used, resulting in a set of KS states that are fractionally occupied.  

While these KS states are formally not physical states, with only the density being physical, their proximity to real, physical states is vital for accurate predictions from the model and is reflected in, for example, Hugoniot predictions \cite{ottoway2021effect, kritcher2020measurement, bethkenhagen2023properties}.  At elevated temperatures therefore, the KS scheme, in which all excited states are represented by a single, averaged excited state, is expected to be a poorer approximation than at zero temperature where, in reality, there is only one occupied state (the ground state) \cite{gonis2018extension}.  This (the single averaged excited state) is also the reason why, for non-degenerate plasmas, the DFT predictions of opacity fail so spectacularly in practice.

A clue to a path forward may be found in the isolated atom approach to opacity \cite{sampson09, fontes2015alamos}.  In that field, an approach known as configuration-average approximation is very similar in practice to the KS-DFT scheme.  One constructs an effective one body interaction potential for the ion or atom, includes an exchange and correlation correction, and solves the equations self-consistently.  The key difference is that one does not use Fermi-Dirac occupations.  Rather, one picks the occupations of the eigenstates to resemble desired excited states.  Instead of there being one, averaged excited state, there are many different excited states, each corresponding to a particular distinct set of chosen occupations.  This approach provides reasonable opacities in non-degenerate plasmas \cite{nagayama19, sampson09, fontes2015alamos}.

However, this isolated atom approach suffers from two major drawbacks when applied to the dense plasma regime: first, it generally ignores, or treats inconsistently, the free electrons; second, it is not variationally derived from an energy expression, and so is generally not reliable for EOS.  A number of works have  improved this situation \cite{gill2023superconfiguration, bar1989super, blenski07, piron13, pain17}, but a practical variational framework has remained elusive.

In this work we give a variationally derived model that contains multiple distinct excited states, and includes free electrons consistently.  As in the isolated atom approach, the excited states are defined by a chosen set of occupations.  The energy of each excited state is calculated from an effective one-particle system with exchange and correlation corrections.  In the limit of one excited state with Fermi-Dirac occupations, this model recovers the KS-DFT.  We give the derivation of this model and apply it to electronic structure and EOS calculations in the dense plasma regime.  For the application considered (excited atoms in a plasma), the model can be viewed as an extension to the work of \cite{liberman,liberman1982inferno}, where the author used DFT to get the averaged properties of an atom in a plasma.  



\section{Free Energy}
We consider an ensemble of nuclei and electrons in a volume $V$ and at temperature $T$.  
In Hartree atomic units, the free energy of the system is
\begin{equation}
\begin{split}
    F = & \sum_x W_x [E_x - T S_x] +T \sum_x W_x \log W_x
\end{split}
\end{equation}
where the sum over $x$ is over non-degenerate (in energy) excited states.
The energy $E_x$ is approximated by
\begin{equation}
    E_x = E_x^{(0)} + E_x^{el} + E_x^{xc} 
\end{equation}
where the effective single particle kinetic energy is 
\begin{equation}
    E_x^{(0)} = \sum_{i \in x} n_{x,i} \int_V d^3r \psi_{x,i}^* (-\frac{1}{2}\nabla^2) \psi_{x,i}
\end{equation}
with $\psi_{x,i}(\br)$ the orbital and $n_{x,i}$ the occupation factor for the $i^{th}$ eigenstate of excited state $x$.
$E_x^{el}$ is the electrostatic energy
\begin{equation}
\begin{split}
    E_x^{el} =& \frac{1}{2} \int_V d^3r \int d^3r' \frac{n_x(\br) n_x(\br')}{|\br -\br'|}\\
            &+ \frac{1}{2}  \sum_m \sum_{n \in V,n\ne m} \frac{Z_n Z_m}{|{\bm R_n} - {\bm R_m}|}\\
            &- \sum_{n \in V} Z_n \int d^3r \frac{n_x(\br)}{|\br - {\bm R_n}|}\\
\end{split}
\end{equation}
with $\bm{R}_i$ the position vector of nucleus $i$, and $n_x(\br)$ the electron density of excited state $x$
\begin{equation}
    n_x(\br) = \sum_{i \in x} n_{x,i}  |\psi_{x,i}(\br)|^2
\end{equation}
and $Z_n$ is the nuclear charge of nucleus $n$.
$E_x^{xc}$ is the exchange and correlation energy, which we approximate with the local density approximation
\begin{equation}
    E_x^{xc} = \int_V d^3r \, \epsilon^{xc}[n_x(\br)]
\end{equation}
The entropy has been split into two contributions (see appendix); a term due to the entropy of the excited state $S_x$
\begin{equation}
\begin{split}
    S_x = &-\int_V d^3r \sum_{i \in x} |\psi_{x,i}(r)|^2 \\
    & \times \left[ n_{x,i} \log n_{x,i} + (1 - n_{x,i}) \log (1 - n_{x,i}) \right] \\
\end{split}
\end{equation}
and a term due to the entropy of mixing $\sum_x W_x \log W_x$, where $W_x$ is the probability of excited state $x$.

\subsection{Constrained Free Energy}
Before minimising this free energy, the following constraints are added
\begin{subequations}
    \begin{equation}
        \sum_{i \in x} \lambda_{x,i} \left[ \int_V d^3r |\psi_{x,i}|^2 -1 \right] = 0\label{eq_c1}
    \end{equation}
    \begin{equation}
        B \sum_x W_x = 1 \label{eq_c2}
    \end{equation}
    \begin{equation}
        \gamma \sum_x W_x \big [\int_V d^3r n_x(\br) - Z_x \big] = 0\label{eq_c3}
    \end{equation}
    \begin{equation}
        \sum_{x,i} \mu_{x,i} W_x \int_V d^3r |\psi_{x,i}(r)|^2 [n_{x,i} - f_{x,i}] = 0\label{eq_c4}
    \end{equation}
\end{subequations}
where $\lambda_{x,i}$, $B$, $\gamma$ and $\mu_{x,i}$ are Lagrange multipliers.
Equation (\ref{eq_c1}) ensures normalization of the orbitals, equation (\ref{eq_c2}) ensures that the probabilities $W_x$ sum to 1, equation (\ref{eq_c3})
requires overall charge neutrality of the excited states, and equation (\ref{eq_c4}) fixes the occupation factors $n_{x,i}$ to be given by the chosen inputs $f_{x,i}$.  This last requirement is how we define an given excited state $x$.

\subsection{Minimization of the Free Energy}
We require the following to be true
\begin{subequations}
\begin{equation}
    \frac{\delta\Omega}{\delta \psi_{x,i}^*(\br)} = 0
\end{equation}
\begin{equation}
    \frac{\partial\Omega}{\partial n_{x,i}} = 0
\end{equation}
\begin{equation}
    \frac{\partial\Omega}{\partial W_x} = 0
\end{equation}
\end{subequations}
where $\Omega$ is the constrained free energy.  Applying the first of these, we obtain
\begin{equation}
\begin{split}
    0 = & W_x \left[ n_{x,i} [-\frac{1}{2}\nabla^2 + V_x^{el}(r) + V_x^{xc}(r)] -  \lambda_{x,i} \right]\psi_{x,i}(r) \\
        & + W_x T \left[ n_{x,i} \log n_{x,i} + (1 - n_{x,i}) \log (1 - n_{x,i}) \right]\psi_{x,i}(r)\\
        & - W_x \gamma n_{x,i} \psi_{x,i}(r)  - W_x \mu_{x,i} n_{x,i} \psi_{x,i}(r)
\end{split}
\end{equation}
which can be rewritten as 
\begin{equation}
\begin{split}
    \left[ -\frac{1}{2}\nabla^2 + V_x^{el}(\br) + V_x^{xc}(\br)  -\gamma \right]\psi_{x,i}(\br) = \epsilon_{x,i} \psi_{x,i}(\br)
    \label{eq_se}
\end{split}
\end{equation}
i.e., the one particle Schr\"odinger equation.  $\gamma$ is determined by setting the zero of the energy.  Further,
\begin{equation}
    V_x^{el} = \frac{\delta E_x^{el}}{\delta n_x(\br)}
\end{equation}
\begin{equation}
    V_x^{xc} = \frac{\delta E_x^{xc}}{\delta n_x(\br)}
\end{equation}

The second minimization requirement gives 
\begin{equation}
\begin{split}
     0 =& W_x\left[ \int d^3r \psi_{x,i}^* (-\frac{1}{2}\nabla^2) \psi_{x,i}\right. \\
        &\left.+\int d^3r |\psi_{x,i}|^2 ( V_x^{el}(\br) +V_x^{xc}(\br) - \gamma - \mu_{x,i})\right] \\
        & + T W_x \int d^3r |\psi_{x,i}|^2 \log \frac{n_{x,i}}{1-n_{x,i}}\\
\end{split}
\end{equation}
which reduces to ($\beta = 1 /T$)
\begin{equation}
\begin{split}
     n_{x,i} = \frac{1}{\exp(\beta (\epsilon_{x,i} - \mu_{x,i})) + 1}
\end{split}
\end{equation}
The $\mu_{x,i}$ are then determined by the requirement that $n_{x,i} = f_{x,i}$, where $f_{x,i}$ is set by input.

The third minimization requirement gives 
\begin{equation}
\begin{split}
     0 = & F_x - B  + T  + T \log W_x \\
\end{split}
\end{equation}
where $F_x = E_x - T S_x $, then
\begin{equation}
\begin{split}
     W_x = & \frac{ \exp\left( -\beta  F_x   \right)} {\cal Z}\\
     \label{eq_wx}
\end{split}
\end{equation}
i.e., the usual Boltzmann factor where
\begin{equation}
\begin{split}
     \cal Z = & \sum_x \exp\left( -\beta  F_x  \right)
\end{split}
\end{equation}
is the partition function.

This completes the model.  First one chooses a set of occupation factors that define the excited states $\{f_{x,i}\}$.  Solving the one particle Schr\"odinger equation (\ref{eq_se}) for each of the excited states can then be carried out in the usual self-consistent field framework.  The excited states are connected though the value of $\gamma$ which can be determined iteratively, though, as we shall see, the EOS does not seem to be sensitive to it for the cases tested here.  Once the energies $E_x$ and entropies $S_x$ of the excited states are determined, the probabilities $W_x$ are found with equation (\ref{eq_wx}).

\subsection{Application to Atomic Model}
Liberman \cite{liberman} introduced the quantum average atom model.  It is a DFT model of an atom in a charge neutral sphere.  The sphere volume is determined by the density of the plasma.  The model uses Kohn-Sham DFT with Fermi-Dirac occupation factors.  Here we extend this concept with the present excited state treatment.  Due to the spherical symmetry of the average, the equations simplify somewhat.  The kinetic energy can be written
\begin{equation}
\begin{split}
        E_x^{(0)} =& \sum_l 2(2l+1) \int d\epsilon\,  n_{x,\epsilon,l} \int_0^R d^3r \\
                   & \times y_{x,\epsilon,l}^* \bigg(-\frac{1}{2}\frac{d^2}{dr^2}+ \frac{l(l+1)}{r^2}\bigg) y_{x,\epsilon,l}
\end{split}
\end{equation}
where
\begin{equation}
\begin{split}
        \psi_{x,i}(\br) = \frac{y_{x,\epsilon,l}(r)}{r}Y_{l,m}(\hat{\br}),
\end{split}
\end{equation}
$y_{x,l,\epsilon}(r)$ is the radial solution \cite{blenski95}, $l$ is the orbital angular momentum quantum number, and the eigenstate index $i$ is replaced with $\epsilon$ and $l$ .  The entropy becomes
\begin{equation}
\begin{split}
    S_x = & -\int_0^R dr \sum_l 2(2l+1) \int d\epsilon  |y_{x,\epsilon,l}(r)|^2 \\
          & \times \left[ n_{x,\epsilon,l} \log n_{x,\epsilon,l} + (1 - n_{x,\epsilon,l}) \log (1 - n_{x,\epsilon,l}) \right]
    \end{split}
\end{equation}
The electrostatic energy is
\begin{equation}
\begin{split}
    E_x^{el} =& \frac{1}{2} \int_V d^3r \int_V d^3r' \frac{n_x(r) n_x(r')}{|\br -\br'|}\\
            &-  Z \int_V d^3r \frac{n_x(r)}{r}\\
\end{split}
\end{equation}
The exchange and correlation term is unchanged.  
The resulting Schr\"odinger equation is
\begin{equation}
\begin{split}
    \left[ -\frac{1}{2}\frac{d^2}{dr^2}+ \frac{l(l+1)}{r^2} + V_x^{el}(r) + V_x^{xc}(r)  -\gamma \right] y_{x,\epsilon,l}(r) \\
    = \epsilon_{x,l} y_{x,l,\epsilon}(r)
\end{split}
\end{equation}
We have brushed over an inconsistency inherent in the average atom models, where the wave function is normalised over all space even though the energy only involves integrals inside the ion sphere.  This issue has been much discussed in the average atom literature and the same problem is inherent in the above atomic model.  It is probably possible to derive a consistent model, as was done for average atoms in Refs. \cite{blenski07,piron11}, but we do not attempt that here.  With this in mind, the pressure can be written 
\begin{equation}
\begin{split}
    P =& -\left. \frac{\partial F}{\partial V} \right|_{T,N} \\
    =&  -\sum_x W_x\left[ f_x(R) \phantom{\sum_x}\right. \\
     & \left. - \sum _{i\in x} \sum_l 2(2l+1) |y_{x,\epsilon,l}(R)|^2 n_{x,\epsilon,l}( \gamma + \mu_{x,\epsilon,l})  \right]
     \label{eq_pressure}
\end{split}
\end{equation}
where $f_x(R)$ is the free energy density evaluated at $R$, and
\begin{equation}
\begin{split}
    \gamma =& \sum_x W_x V_x^{xc}(R)\label{gamma}
\end{split}
\end{equation}

\section{Numerical Results}
\subsection{Choosing the occupation factors}
We now apply this atomic model to the calculation of EOS in dense plasmas.  The first step is to enumerate the states so that a list of all permutations can be created.  There are an infinite set of possible excited states that one could consider based on variations of the occupation factors of bound and continuum electrons. With this in mind, we create an approximate, coarse enumeration of the states by defining energy boundaries $\epsilon_x^{n,l}$
\begin{equation}
    2(2l+1) = \int_{\epsilon_x^{n,l}}^{\epsilon_x^{n+1,l}}d\epsilon\, \chi_{x,l}(\epsilon)\label{eq_ei_def}
\end{equation}
\begin{figure}
\begin{center}
\includegraphics[scale=1]{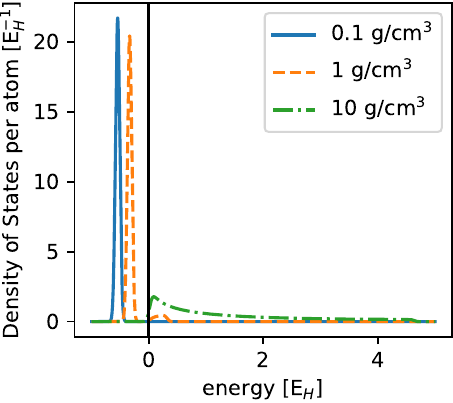}
\end{center}
\caption{Filled density of states for helium.  We have applied an arbitrary broadening of 1 eV to the DOS so that the bound states, which would otherwise be delta functions, can be displayed. }
\label{fig_he_dos}
\end{figure}
where $n = 1,2,\ldots$  The first energy bound for a given $l$, $\epsilon_x^{1,l}$ is chosen to be just lower in energy than the lowest energy eigenstate, for that $l$, and $\chi_{x,l}$ is the density of states for excited state $x$ for angular momentum  quantum number $l$,
\begin{equation}
     \chi_{x,l}(\epsilon) = 2(2l+1) \int_V d^3r |y_{x,\epsilon,l}(r)|^2.
\end{equation}
With the set of states defined by these energy ranges, we choose to occupy them with all perturbations of integer occupations as is common in atomic physics approaches.  We must also choose a maximum $n$ to consider.  The quantity $n$ becomes equivalent to the principle quantum number for bound states that are well contained inside the ion-sphere.  Hence, for example, if we have a helium plasma, and consider the maximum $n$ to be 2 ($n_{max} = 2$), we would have the following list of excited states:
\begin{enumerate}
    \item $1s^2\, 2s^0 \, 2p^0$
    \item $1s^1\, 2s^1 \, 2p^0$
    \item $1s^1\, 2s^0 \, 2p^1$
    \item $1s^0\, 2s^2 \, 2p^0$
    \item $1s^0\, 2s^1 \, 2p^1$
    \item $1s^0\, 2s^0 \, 2p^2$
\end{enumerate}
and for configuration 2, for example, we have $n_{2,\epsilon_{1s},0} = 0.5$.

In Figure \ref{fig_he_dos} we show the filled DOS for this example for configuration 1.  At low densities the bound state is well contained within the ion sphere and there are no free electrons.  As density increases, the $1s$ orbital becomes partially bound, and all two electrons cannot fit into the bound state, so the upper energy limit of this state extends into the continuum, satisfying the definition (\ref{eq_ei_def}).  For the highest density, there are no bound states and the $1s$ is extended into the continuum.  We see that this atomic picture of excitations will be best if there are no (or very few) free electrons in the plasma since, in our example, for 10 g/cm$^3$, the two ionized electrons are spread evenly over the $1s$ band, which is only one possible distribution out of infinitely many.

There is another problem with this above approach to excited states -- it will be prohibitively expensive for high temperatures, or higher $Z$ materials, where $n_{max}$ must be large.  A solution is to treat all states with energies higher than $\epsilon_{x,\epsilon_x^{n_{max}+1},l}$ with Fermi-Dirac statistics.
The lifetime of excitations of core (deeply bound) states is relatively long, and that for excitations of continuum (free) electrons is much shorter due to collisions. Therefore, it is reasonable to create a set of occupation factors that only include a detailed list of excitations for the core states, and use an average occupation (Fermi-Dirac) for the free electrons, at least for EOS purposes. In the results presented here we have used this approach.  For our helium example, the list of excited states increases to include
\begin{enumerate}
\setcounter{enumi}{6}
    \item $1s^1\, 2s^0 \, 2p^0$ + FD
    \item $1s^0\, 2s^1 \, 2p^0$ + FD
    \item $1s^0\, 2s^0 \, 2p^1$ + FD
    \item $1s^0\, 2s^0 \, 2p^0$ + FD
\end{enumerate}
where the occupation factors are Fermi-Dirac beyond last energy boundary for each $l$ runs, in principle, to $\infty$.

We note that by using Fermi-Dirac occupation factors in the definition of an excited state, the configurations become dependent on temperature, whereas in the first example, the electronic structure of the configurations and their energies are independent of temperature.  For configurations that are independent of temperature, the resulting probabilities (populations) do depend on temperature, but simply, through the free energy ($F_x = E_x - T S_x$) and equation (\ref{eq_wx}).

\subsection{Application to Aluminum Plasmas}
\begin{figure}
\begin{center}
\includegraphics[scale=1]{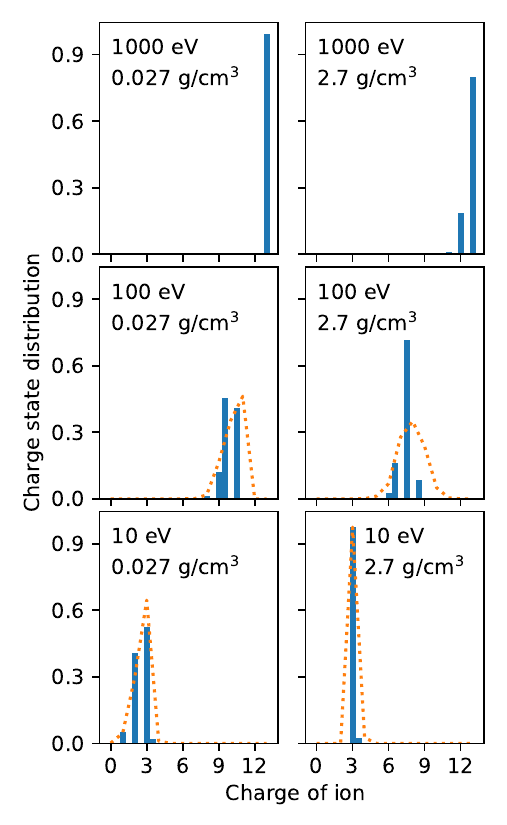}
\end{center}
\caption{Charge state distributions for aluminum plasmas.  We have considered configuration perturbations up to the $n=2$ shell for 2.7 g/cm$^3$, and $n=3$ shell for 0.027 g/cm$^3$.  The results of the present model are shown in the blue bars.  Also shown with the dotted lines are the results of reference \cite{white2022charge}. }
\label{fig_al_csd}
\end{figure}
In Figure \ref{fig_al_csd} the charge state distribution (CSD) for aluminum plasmas is shown.  Here we calculate the charge of an ion by counting the number of positive energy electrons, which therefore includes electrons in resonance states.  This is a reasonable definition, but we note that with this definition it is possible to have an ion of non-integer charge.  This can been seen from Figure \ref{fig_he_dos} where a bound state is partially bound for 1 g/cm$^3$.  Physically, this behavior reflects the fact that as a bound state pressure ionizes it is neither truly a bound nor free electron state.  In Figure~\ref{fig_al_csd} we see that at the highest temperature (1000 eV) and lowest density (0.027 g/cm$^3$), the plasma is fully ionized.  Going next to the 1000 eV, 2.7 g/cm$^3$ case, we find that the plasma is mostly fully ionized but contains about 20\% of ions with a single bound electron.  Increased collisional effects at this higher density are the cause of the lower average ionization.  
\begin{figure}
\begin{center}
\includegraphics[scale=1]{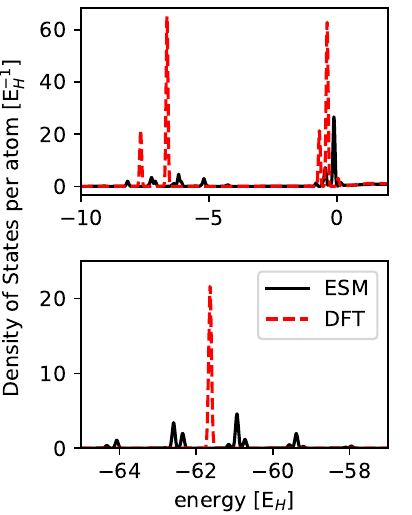}
\end{center}
\caption{Filled density of states for aluminium at 100 eV and 2.7 g/cm$^3$.  The solid black line is the result from the present excited states model (ESM).  This is compared to the DFT prediction in dashed red from the average atom model \texttt{Tartarus} \cite{starrett19}.  }
\label{fig_al_dps}
\end{figure}

For the 10 and 100 eV cases we compare to the model of White \emph{et al.} \cite{white2022charge}, Figure \ref{fig_al_csd}.  For the 100 eV, 2.7 g/cm$^3$ case the new model is significantly more strongly peaked.  For the 2.7 g/cm$^3$ cases we have used $n_{max} = 2$.  In the limit of using $n_{max} = 0$ we would recover the DFT result and the CSD would be peaked at one charge state with all of the population.  Using $n_{max}=1$ would allow some fluctuation, $n_{max}=2$ even more, and so on.  Hence, the reason why the CSD of the new model is more peaked than that of White et al \cite{white2022charge} is due to the lower value of $n_{max}$.  Does this mean that the result of the present model for this case is not converged with respect to $n_{max}$?  As we argued earlier, the present definition of the excited state works best for core states.  Using $n_{max}=2$ in this case ensures this outcome, while using $n_{max}=3$ would not as the $n=3$ states are partially bound.  Which choice is more reasonable depends on the time scale of the experiment that we wish to model.  Excitations of the partially bound $n=3$ states will have a much shorter lifetime than those of the core states.  For experiments that  probe time-integrated quantities over time-scales that are longer than the $n=3$ excitation lifetimes, the present model is appropriate, whereas, if the experiment has shorter time resolution then considering explicit excitations of these shells would be necessary.  

Lastly, we note that the CSD is an output of the model and we are free to choose different definitions of ion charge that are reasonable.  This choice does not affect the model in any way -- it does not change the EOS, energies, entropy, nor populations.

In Figure \ref{fig_al_dps} the filled density of states is shown for a hot dense aluminum plasma at 100 eV and 2.7 g/cm$^3$ using $n_{max}=2$.  With aluminum's 13 electrons this leads to 63 distinct excited states and the result shown is averaged over all of these with the probabilities $W_x$.  In the lower panel we see the effect on the $1s$ eigenstate.  There are four distinct groupings of $1s$ states.  These correspond to four different charge states with significant probability in the plasma.  In contrast, we also show the DFT result obtained using the \texttt{Tartarus} code.  There is only one peak, corresponding to the average $1s$ energy of the plasma.  The different peaks within each cluster correspond to different arrangements of electrons in the `spectator' bound states (i.e., those in the $n=2$ shell), but having the same ion charge.

In the top panel of Figure \ref{fig_al_dps} we again see a cluster of peaks from -10 to -5 E$_H$ corresponding to the $n=2$ shell.  The DFT results show two distinct peaks corresponding to the $2s$ and $2p$ eigenstates.  Near zero energy, the line corresponding to the $n=3$ states shows up.  Since they are not explicitly included in our excited state list, the excited state model predicts a strong line near zero energy.

\begin{figure}
\begin{center}
\includegraphics[scale=1]{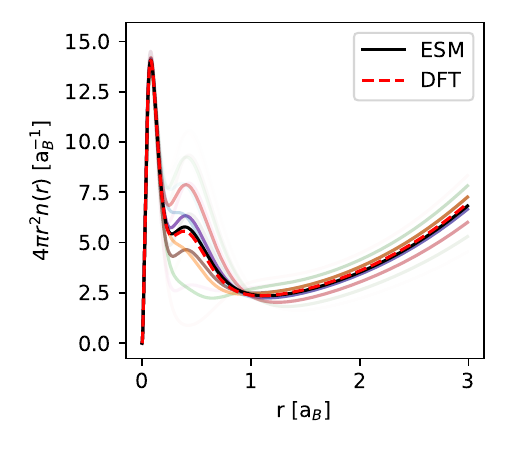}
\end{center}
\caption{Electron density for aluminium at 100 eV and 2.7 g/cm$^3$.  The solid black line is the predicted average radial density from the present excited states model (ESM).  This is compared to the DFT prediction in dashed red from the average atom model \texttt{Tartarus} \cite{starrett19}.  Also shown in semi-transparent lines are the individual densities due to the excited states, with their degree of transparency proportional to to the probability $W_x$.}
\label{fig_al_density}
\end{figure}

In Figure \ref{fig_al_density} we show the radial density averaged over all excited states
\begin{equation}
    n(\br) = \sum_x W_x n_x(\br)
\end{equation}
We compare this to the DFT result from the \texttt{Tartarus} model.  We find overall good agreement, but some differences are observed.  On the one hand, it is not surprising that the average over the excited states is not the same as the averaged excited state. On the other, the difference is fairly small and would be hard to test experimentally.  Also shown in the figure are the electron densities from the excited states.  These curves indicate that there is little fluctuation in the occupation of $1s$ shell (the peak nearest the origin), there is significant variation on the $n=2$ shell occupation (the second peak from the origin), and significant variation in the free electron density (the tail after the peaks).   We note that while the DOS, Figure \ref{fig_al_dps}, indicates that the $1s$ eigenvalue does depend on the charge state, Figure \ref{fig_al_density} shows that these variations do not strongly affect the density due to the $1s$ shell.

In Table \ref{tab_pressure} the excess pressure is shown (given by equation (\ref{eq_pressure}) which does not include the ideal ion contribution) for the same aluminum plasmas, ranging from degenerate systems (2.7 g/cm$^3$ and 10 eV) to fully ionized (0.027 g/cm$^3$ and 1000 eV).  These pressures are compared to \texttt{Tartarus} results.  Overall agreement is close between the two methods.  Agreement is best for the high temperature cases, where details of the interaction potentials are unimportant due to the high average energy of the free electrons, as well as for the most degenerate case, where the plasma is dominated by one charge state (i.e., close to the DFT limit of the model).  The remaining differences are relatively small and can be explained by the difference between calculating the pressure of an average system (DFT) versus the average pressure of the resolved excited states.
\begin{table}
\begin{center}
\bgroup
\def\arraystretch{1.5}%
\begin{ruledtabular}
\begin{tabular}{cc cccccc}
 &Al 0.027 g/cm$^3$& \\
 & ESM (n=3) & \texttt{Tartarus}\\
10 eV & 2.02e-2 & 2.16e-2 \\ 
100 eV  & 0.954   & 0.948 \\ 
1000 eV  & 12.5 &  12.5 \\[0.3cm]\hline
&Al 2.7 g/cm$^3$& \\
 & ESM (n=2) & \texttt{Tartarus} \\
10 eV & 2.08 & 2.07 \\
100 eV & 64.4 & 65.4 \\
1000 eV & 1203 & 1201\\
\end{tabular}
\end{ruledtabular}
\egroup
\end{center}
\caption{Values of excess pressure in Mbar for aluminum at temperatures of 10, 100 and 1000 eV and densities of 0.027 g/cm$^3$ and 2.7 g/cm$^3$.  ESM refers to the present excited states model, while \texttt{Tartarus} refers to the DFT based average atom model \cite{starrett19}.\label{tab_pressure}}
\end{table}

The calculation of $\gamma$ (equation \ref{gamma}) requires an initial guess and an iterative, self-consistent, procedure.  We start with the value provided by an average atom model which seems to be close to the final answer.  For example, for Al at 100 eV and 2.7 g/cm$^3$, the value from the \texttt{Tartarus} model is $\gamma_{AA} = -0.446$ E$_H$ (which results in a pressure of 64.44 Mbar), while the converged value is  $\gamma = -0.444$ E$_H$ (which results in a pressure of 64.41 Mbar).  Conveniently, an approximation in which we choose $\gamma$ to vary for each excited state such that $\gamma_x = V_x^{xc}(R)$, and using $\gamma = \sum_x W_x \gamma_x$ also seems to be accurate for EOS, giving $\gamma = -0.444$ E$_H$ and a pressure of 64.43 Mbar for this case.  Clearly, this approximation could be more problematic for excitation energies (differences in excited state energies) than for the EOS which is a more averaged quantity.

\section{Conclusions}
A variational model of excited states in electronic structure has been presented.  The model recovers the usual Kohn-Sham density functional theory approach in the limit where only one state dominates (i.e., for degenerate systems like solids, or for fully ionized plasmas).  The model uses an effective one-electron expression for the excited state energy and includes the LDA for the exchange and correlation energy.  Boltzmann factors for the excited state probabilities result from minimising the free energy with respect to the probabilities.

Excited states are defined by a set of one-electron level occupation factors.  If these are set to be the Fermi-Dirac occupation factors, then Kohn-Sham DFT is recovered.  We apply this variational theory to a model of an atom in a plasma; a generalization of the average atom model \cite{liberman, liberman1982inferno}.  We use an atomic physics-inspired definition of excited states, where permutations of integer occupations of bound states are considered.  Comparison of this application to the average atom model \texttt{Tartarus} is made.  We see the effect on the density of states, density and pressure.  In general the pressure is quite close to the DFT calculation but some differences are observed.  Since DFT is a widely used and trusted method, this can be considered as a validation of the current model.  The advantage of the current approach is that the calculation of excited states should allow prediction of more realistic optical properties with a consistent and realistic EOS.  This advantage remains to be explored.

\section*{Acknowledgments}
LANL is operated by Triad National Security, LLC, for the National Nuclear Security Administration of the U.S. Department of Energy under Contract No.~89233218NCA000001.

\appendix

\section{Entropy}
The entropy is given by
\begin{equation}
    S = - \sum_i W_i \ln W_i
\end{equation}
where the sum is over all microstates of the system.  If microstate $i$ is degenerate, with $g_x$ being the total number of microstates at the energy $E_x$, then
\begin{equation}
    S = - \sum_x \sum_{i \in x} W_i \ln W_i
\end{equation}
the probability of microstate $i$ is $W_i = W_x / g_x$, so
\begin{equation}
\begin{split}
    S = & - \sum_x \sum_{i \in x} \frac{W_x}{g_x} \ln \frac{W_x}{g_x} \\
     = &- \sum_x W_x \ln \frac{W_x}{g_x}\\
     = &- \sum_x W_x  \ln W_x +\sum_x W_x S_x \\
\end{split}
\end{equation}

\vspace{0.3in}

\bibliographystyle{unsrt}
\bibliography{phys_bib}

\end{document}